\documentclass[pre,twocolumn]{revtex4}
\usepackage{amsmath,graphicx}

\def\aver#1{\left\langle#1\right\rangle}

\begin{document}

\title{Long-lived states in synchronized traffic flow.\\
Empirical prompt and dynamical trap model}

\author{Ihor Lubashevsky}
\affiliation{Theory Department, General Physics Institute, Russian Academy of
Sciences, Vavilov str. 38, Moscow, 119991, Russia}
\author{Reinhard Mahnke}
\affiliation{Fachbereich Physik, Universit\"at Rostock, D--18051 Rostock,
Germany}
\author{Peter Wagner}
\affiliation{Institute of Transport Research, German Aerospace Center (DLR),
Rutherfordstrasse 2, 12489 Berlin, Germany.}
\author{Sergey Kalenkov}
\affiliation{Physics Department, Moscow State University for Technology
``MAMI'', Bol'shaya Semenovskaya str.~38, Moscow, 105831, Russia}
\date{\today }

\begin{abstract}
The present paper proposes a novel interpretation of the widely
scattered states (called synchronized traffic) stimulated by Kerner's
hypotheses about the existence of a multitude of metastable states in
the fundamental diagram. Using single vehicle data collected at the
German highway A1, temporal velocity patterns have been analyzed to
show a collection of certain fragments with approximately constant
velocities and sharp jumps between them. The particular velocity values
in these fragments vary in a wide range. In contrast, the flow rate is
more or less constant because its fluctuations are mainly due to the
discreteness of traffic flow.

Subsequently, we develop a model for synchronized traffic that can
explain these characteristics. Following previous work
(I.~A.~Lubashevsky, R.~Mahnke, Phys.~Rev.~E {\bf 62}, 6082, 2000) the
vehicle flow is specified by car density, mean velocity, and
additional order parameters $h$ and $a$ that are due to the
many-particle effects of the vehicle interaction. The parameter $h$
describes the multilane correlations in the vehicle motion. Together
with the car density it determines directly the mean velocity. The
parameter $a$, in contrast, controls the evolution of $h$ only. The
model assumes that $a$ fluctuates randomly around the value
corresponding to the car configuration optimal for lane changing. When
it deviates from this value the lane change is depressed for all cars
forming a local cluster.  Since exactly the overtaking manoeuvres of
these cars cause the order parameter $a$ to vary, the evolution of the
car arrangement becomes frozen for a certain time. In other words, the
evolution equations form certain dynamical traps responsible for the
long-time correlations in the synchronized mode.
\end{abstract}

\maketitle

\section{Anomalous property of the synchronized mode}

Although the motion of individual vehicles is controlled by the
motivated driver behavior rather than by simple physical laws the car
ensembles on highways exhibit phenomena widely met in physical
systems. Namely, the existence of various states (self-sustained
steady-state modes of traffic flow), their coexistence, the phase
transitions, \textit{etc.} (reviewed in \cite{KTGF99,KIsr,KL3}),
actually forms a novel branch of physics.  Following the pioneering
works by Lighthill \& Whitham~\cite{LW}, and Richards~\cite{R} the
state of traffic flow on a highway is specified by the vehicle density
$\rho $ and the mean velocity $v$ or, equivalently, by the traffic
flow rate $q=\rho v$. In other words, the vehicle density $\rho$ and
the flow rate $q$ are regarded as the complete set of state variables,
leading to a correspondence between a point in the $\rho q$-plane
with a particular state of traffic flow. The main assumption adopted
in the classical theories (reviews can be found in
\cite{Nag96,Helbook,Sh00,Hrev,Hrev2}) is the existence of a
relationship between the mean velocity and vehicle density,
$v=\vartheta (\rho )$. So, the curve $q=\rho \vartheta (\rho )$ in the
$\rho q$-plane is called fundamental diagram which characterizes the
possible states of traffic flow.

\begin{figure}
\begin{center}
\includegraphics{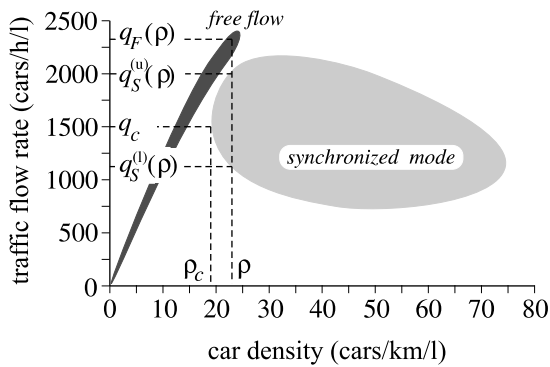}
\end{center}
\caption{Illustration of the experimental data by~\protect\cite{K2}
  showing possible states of the free flow and the synchronized mode
  in the $\rho q$-plane.\label{exp}}
\end{figure}

According to traffic flow data analyzed recently by Kerner \& Rehborn
\cite{K1,K2,K3}, Kerner~\cite{K4,K5}, and also Neubert \textit{et al.}
\cite{Neub99} there are three distinctive states of multilane traffic
flow, the free flow ($F$), the synchronized mode ($S$), and wide
moving jams ($J$). This endows the multilane traffic flow with a
variety of properties. In particular, it has been demonstrated that
the self-formation of moving jams mainly proceeds via the sequence of
two phase transitions $F\rightarrow S\rightarrow J$ \cite{K4}. Both of
them are of the first order, i.e.\ exhibiting breakdown, hysteresis,
and nucleation effects \cite{K3}. The present paper focuses its
attention on a unique property of the synchronized mode itself, its
complexity \cite{K2}. In contrast to the free flow, the synchronized
mode matches a two-dimensional domain on the $\rho q$-plane rather
than a curve on it (Fig.~\ref{exp}). It means that a track made up of
the empirical points $\{\rho(t),q(t)\}$ obtained at sequential time
moments unpredictably fills a two-dimensional domain. In this regard
the synchronized mode is also referred to as a widely scattered
traffic state for which a fundamental diagram in the form of a
one-dimensional curve did not exist.

The synchronized mode is characterized by strong correlations in the
vehicle motion at different lanes ordering the car arrangement across
the road. To describe this state a model has to take into account
particular details of the multilane car interaction. Whence it follows
that the vehicle density $\rho$ and the mean velocity $v$ averaged
over all the lanes across the highway form an incomplete set of state
variables. One of the possible ways to tackle this problem is to
ascribe to each lane its own vehicle density and mean velocity.
Several models for multilane traffic flow applying to such an approach
have been proposed, in particular, gas-kinetic
theory~\cite{GK1,GK2,GK3,GK4,GK5,GK6} and compressible fluid
model~\cite{FL}. For a more detailed survey, including also the
cellular automata multilane models, we refer to
\cite{Nag96,Helbook,Sh00,Hrev,Hrev2}.  Here we only note that in all
such models the cause of the free flow instability is related to a
delay in the driver's response to variations of the headway distance.
This instability gives rise to the traffic flow pinch leading to the
formation of a new congested phase of vehicle motion.

In spite of the variety of available models for multilane traffic flow
up to now no generally accepted explanation for the widely scattered
states of the synchronized mode has been proposed. There are various
points of view on the mechanism responsible for this phenomenon. In
particular, it is a mixture of different vehicle types (cars and
trucks)~\cite{17,NN1}, a heterogeneity in the headway
distance~\cite{18} as well as in the highway structure~\cite{18a},
changes in the behavior of ``frustrated'' drivers~\cite{19},
anti\-cipation effects~\cite{20,21,22,23}, non-unique equilibrium
solutions of the Prigogine-Herman kinetic equation~\cite{24}, the
existence of certain plateaus in the dependence of the optimal
velocity on the headway distance~\cite{NN2,Peter1}, and the variety of
oscillating metastable states with different wavelengths~\cite{25,26}.
The requirement of multilane models to describe synchronized traffic
flow has already been stated in \cite{GK3,LLK98} using hydrodynamic as
well as kinetic equations.

Commenting on these models we would like to say the following. First,
most of them apply to a one-dimensional representation of traffic
flow, discrete or continuum one. Therefore they do not take into
account the strong multilane vehicle interaction which is an essential
feature of the synchronized mode. Such a theory can explain the
complex behavior of traffic flow on a single-lane road rather than on
multilane highways or can describe the congested multilane traffic
flow providing that the multilane synchronization is a minor effect.
Second, according to the analysis of single-vehicle data
in~\cite{Neub99} the synchronized mode is mainly singled out by strong
multilane correlations in the vehicle velocity rather than the formal
equality of speed at which cars moving on different lanes. In
particular, in the observed free flow regime the averaged velocities
at various lanes differ from one another only slightly. Therefore the
approach based on the description of multilane traffic flow in terms
of individual streams along different lanes with their own mean
velocities and vehicle densities seems to be doubtful.

Following the classification proposed by Kerner \& Rehborn \cite{K2}
there are three distinctive kinds of synchronized flow: (\textit{i})
stationary and homogeneous states where both the average speed and
flow rate are approximately constant during a fairly long time
interval, (\textit{ii}) states where only the average vehicle speed is
stationary (`homogeneous-in-speed-states'), and (\textit{iii})
non-stationary and non-homogeneous states. It should be noted that on
macroscopic scales congested traffic exhibits a large variety of
different phenomena, causing a more detailed classification
\cite{L2,L3,Ha1,Ha2} (see also review \cite{Hrev}). Recently, Kerner
formulated several hypotheses about the synchronized mode
\cite{KTGF99,KIsr,KL3,K2,H1H2} to explain the properties of the widely
scattered states. In particular, regarding the synchronized traffic of
types (\textit{i}) and (\textit{ii}) he assumed it to contain a whole
multitude (continuum) of possible congested traffic states
\cite{K2,H1H2}, which are stable with respect to infinitesimal
perturbations. This leads to continuous spatial-temporal transitions
between these states~\cite{KTGF99,KIsr,KL3,H1H2}. However, simulation
models and further empirical evidence for these hypotheses are still
to be found~\cite{Hrev}.

\begin{figure*}
\begin{center}
\includegraphics{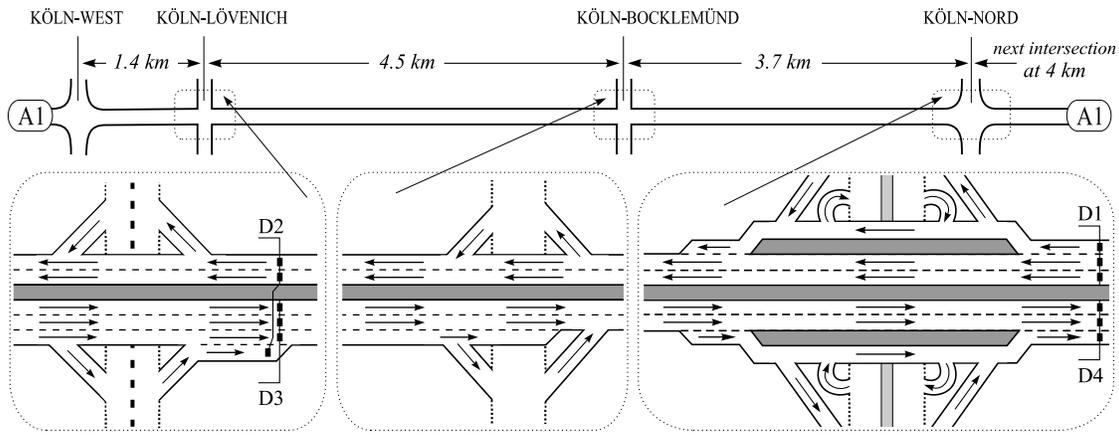}
\end{center}
\caption{ The analyzed section of German highway A1 and position of the loop
  detectors D1 -- D4. Arrows indicate the direction of vehicle
  motion.\label{roadsketch}}
\end{figure*}

According to the prevailing notion (see, e.g.\ Ref.~\onlinecite{Hrev})
the synchronized mode is mainly identified by the absence of a direct
relationship between the locally averaged vehicle velocity and density
provided the case of the ``stop-and-go wave'' pattern has been
excluded from consideration.  Therefore, the main part of the
theoretical models relates the synchronized mode to the instability of
the homogeneous traffic flow caused by the driver delay in the
response to variations in the headway distance. These models explain
the existence of the widely scattered traffic states taking into
account the complex behavior of the developed spatial structure. From
our point of view this is a quite natural approach that describes only
the synchronized mode of type (\textit{iii}), called ``oscillating
congested traffic'' \cite{Hrev}. However, to justify Kerner's
hypotheses concerning types (\textit{i}) and (\textit{ii}) in the
framework of such an approach a traffic flow model has to admit a
large variety of stable nonhomogeneous solutions that either are fixed
in the space or move as a whole with a constant speed. Such a model
has been proposed in paper~\cite{25} based on the car-following
approach where the existence of the variety of stable nonhomogeneous
solutions differing in wavelength has been found numerically. It is an
open question whether this variety is due to the periodic boundary
conditions adopted in the model or is an inherent property. At least
under random noise, the model demonstrates a certain selection of a
single stable solution, so, it seems to describe actually type
(\textit{iii}) of the synchronized mode or a certain mixture of types
(\textit{ii}) and (\textit{iii}). We would like to underline once more
that the cause of such a congested traffic state is related to the
delay in the driver response giving rise to the instability of
homogeneous traffic flow and the multilane synchronization is a
secondary effect only. Besides we note paper~\cite{26} that models the
widely scattered state region by applying actually the idea used also
by Kerner~\cite{KTGF99,KIsr,KL3,H1H2} in justifying his hypotheses.
Namely it is assumed that in congested traffic drivers try to maintain
a constant time to collision. The latter is possible within a wide
interval of headway distances if the velocity difference between cars
is sufficiently small. However this effect is a consequence of a
certain degeneration of the adopted model. Introducing other terms
they can again produce the oscillating behavior of the developed
nonhomogeneous solution.

This paper presents a new approach to describe the widely scattered states for
the synchronized mode of types (\textit{i}) and (\textit{ii}) where the
multilane car correlations are fundamental in understanding the origin of
congestion. The notion `synchronized' traffic seems to embrace distinguished
modes of congested multilane traffic flow. So we just deal with another type
of the synchronized mode. We think that the role of the multilane correlations
in the ``oscillating congested traffic'' is either of minor importance or is
reduced to prevent the traffic flow going into the jammed ``stop-and-go way''
state. In the next section we will try to show that there is another type of
synchronized mode further called `light' synchronized traffic that meets the
aforementioned Kerner's hypotheses. Namely, in this mode the traffic flow
dynamics contains fragments at which the vehicle velocity keeps its value
approximately unchanged inside relatively long time intervals. The transitions
between these fragments proceed through sharp jumps and the particular values
of such a quasi-stationary vehicle velocity vary inside a wide interval. The
subsequent sections will present the corresponding model developed from our
approach formulated in papers~\cite{W1,W2}. It should be noted that in the
presented model these quasi-stationary states are not stationary at all, at
least in the standard physical meaning, so we prefer to call them long-lived
states of synchronized traffic.

\section{Single-vehicle data and two types of the observed
synchronized traffic}

In this section we analyze single-vehicle data collected by loop
detectors at the German highway A1 near Cologne between 6-th and 17-th
June 1996.  Figure~\ref{roadsketch} illustrates the analyzed section
of the highway and the position of the detectors. A detailed
description of this data set and the detection devices has been
presented in~\cite{Neub99}. Here, we only briefly recall their main
features. The two sets of detectors D1 and D4 were placed nearby the
busy intersection between the highway A1 and the highway A 57
(K\"oln-Nord), while the detectors D2 and D3 are located close to the
on-ramps and exit-ramps of the junction K\"oln-L\"ovenich.  In between
there is another junction (AS K\"oln-Bocklem\"und) but with a rather
low usage. The most pronounced congested traffic was observed at the
detector D1 where the number of lanes is reduced from three to two for
cars passing the intersection towards K\"oln-L\"ovenich. Therefore,
the data collected by the selected detector were mainly studied
in~\cite{Neub99}. The congested traffic observed near the detector D1
is characterized by a sharp fall in the mean velocity persisting
within several hours. In connection with the following we prefer to
call it heavy synchronized mode or ``oscillating congested traffic''.

The detectors D2, D3, and D4 also recorded congested traffic that,
however, seems to be of another nature that will be called light
synchronized mode in the following. Exactly in the vicinity of the
detectors D2, D3, and D4 traffic flow demonstrated the behavior
meeting Kerner's hypotheses in the sense discussed at the end of the
previous section. The present paper does not pretend to a thorough
analysis of these data, we only demonstrate the fact that the
hypothetical behavior of traffic flow has been recorded directly and
is likely not to be a misinterpretation of non-stationary
inhomogeneous vehicle structures moving on the highway.

For each detector the data set is mainly composed of the sequence of
numbers $\{t_i,v_i,l_i,k_i\}$ showing that $i$-th car of type $k_i$
(passenger car, truck, truck trailer \textit{etc}.) passed the given
detector at time $t_i$ moving on lane $l_i$ with speed $v_i$. It
should be pointed out that the passing times $\{t_i\}$ were measured
with second's accuracy. Below, the time variations of $q_l(t,\Theta)$
and $v_l(t,\Theta)$ will be analyzed. Here, $q_l(t,\Theta)$ is the
traffic flow rate and $v_l(t,\Theta)$ the vehicle velocity on lane $l$
at time $t$ (measured in seconds) averaged over the time interval
$2\Theta$ according to:
\begin{eqnarray}
  \label{qdef}
  q_l(t,\Theta) & = & \sum_{t_i} w_{\Theta}(t-t_i)\,,\\
  \label{vdef}
  v_l(t,\Theta) & = & \frac{1}{q_l(t,\Theta)}\sum_{t_i} w_{\Theta}(t-t_i)v_i\,.
\end{eqnarray}
The weight coefficients $w_{\Theta}(t)$ are given by:
\begin{equation}\label{wcoef}
  w_{\Theta}(t) =
\begin{cases}
  Z\exp\left(-\left|t\right|/\Theta\right)
  &
  \text{for $\left|t\right|\leq 2\Theta$,}\\
  0 &
  \text{for $\left|t\right| > 2\Theta$,}
\end{cases}
\end{equation}
subject to the normalization condition
\begin{equation}\label{ncond}
  \sum_{t=-\infty}^\infty w_{\Theta}(t) = 1,
\end{equation}
For $\Theta \gg 1$, the factor $Z$ is given by
$Z \simeq [2\Theta(1-e^{-2})]^{-1}$. According to this expression the
characteristic time scale of averaging with the weight $w_{\Theta}(t)$
is $2\Theta$, that is why we will refer to the time interval of
averaging the data set keeping in mind the value $2\Theta$ rather than
$\Theta$.

\begin{figure*}[!]
\begin{center}
\includegraphics[scale=1]{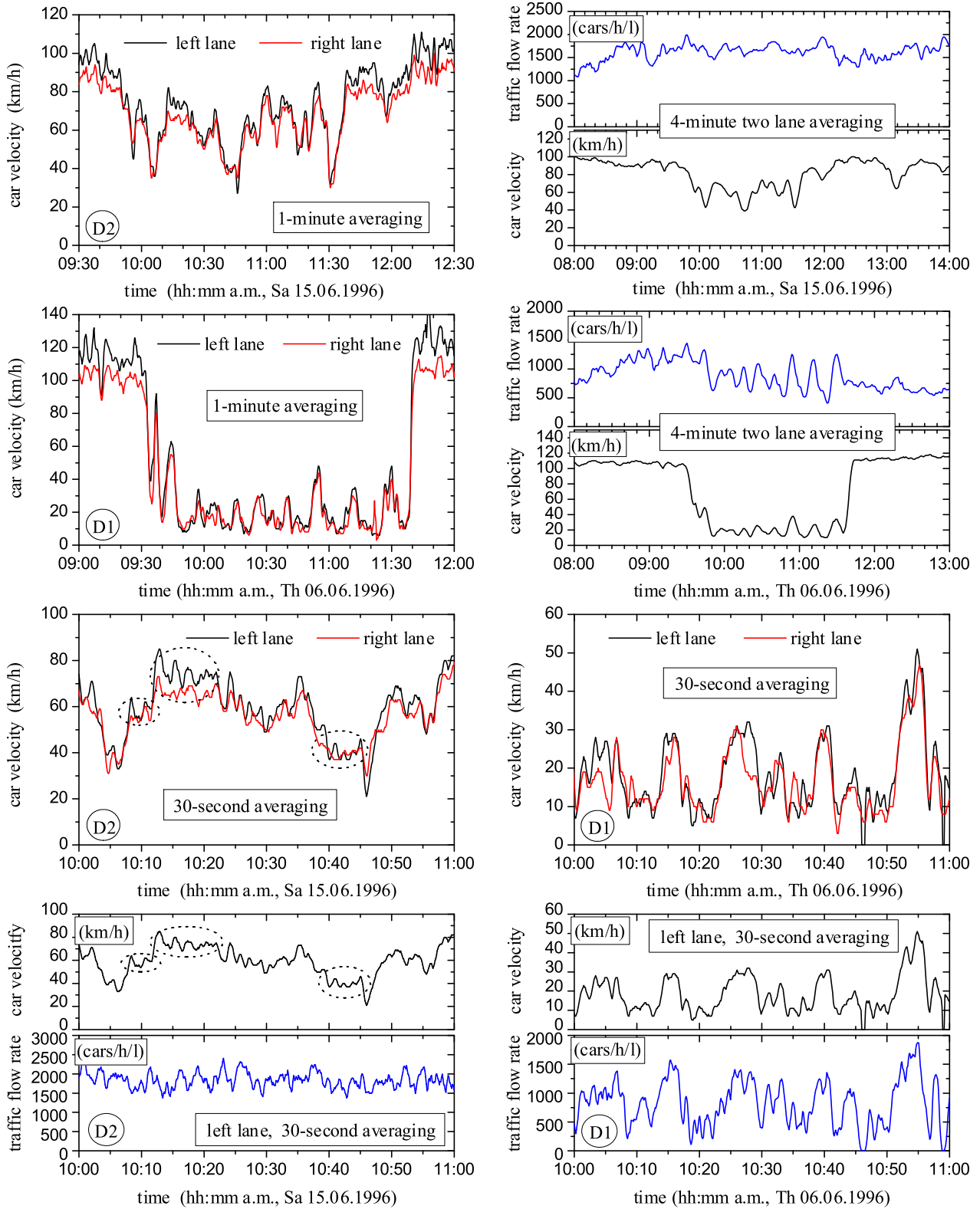}
\end{center}
\caption{%
  Time variations in the vehicle velocity and flow rate observed at
  the detectors D1 and D2 on Th 6 June 1996 and Sa 15 June 1996,
  respectively. This figure visualizes the characteristic properties
  of two limit types of the synchronized mode, lightly congested
  traffic (D2) and heavily (D1) one.
  Red color is chosen to lighten
  the comparison of the vehicle velocity dynamics at the left and
  right lanes, the blue curves are the flow rate dynamics.
  Dotted
  ellipses single out possible fragments of time variations in the
  vehicle velocity that can be referred to as the long-lived states of
  the synchronized mode. In the text particular windows are labeled
  with the row number counted from the top and the side (left or
  right). \label{fignew1}}
\end{figure*}

As has been mentioned above, heavy congested traffic is a typical case
in the vicinity of the detector D1 during the morning hours of
workdays and is accompanied with a substantial fall in the mean
vehicle velocity. The other detectors measured solely light congested
traffic when the mean velocity, on the average, does not drop
essentially but exhibits strong fluctuations only. To illustrate these
types of traffic flow we have chosen, by way of example, the time
series of velocity and flow collected by the detectors D1 and D2 on Th
6 June 1996 and Sa 15 June 1996, respectively. Figure~\ref{fignew1}
visualizes these data. The row~1 left and row~2 left windows show time
variations in the vehicle velocities at different lanes averaged over
1~minute time interval according to Eq.~(\ref{vdef}). The
visualization time intervals (from 9:30 to 12:30 a.m. for D2 and from
9:00 to 12:00 a.m. for D1) span all the time during which the
congested flow was recorded on these days. In spite of the clearly
visible difference in the time patterns at the detectors D2 and D1 the
vehicle velocities at the left and right lanes are strongly
correlated. So both the observed modes of congested traffic can be
categorized as the synchronized traffic flow.  It should be noted that
the mean vehicle speed at the off-ramp lane near the detector D1
(Fig.~\ref{roadsketch}) was at least twice as high as the mean speeds
at the inner-lanes on 6 June 1996 and did not exhibit strong
fluctuations during the given time interval. It seems that the traffic
flow on this off-ramp was ``free'' in comparison with the main stream
and did not affect it essentially.

The row~1 right and row~2 right windows demonstrate time variations in
the vehicle velocity and the traffic flow rate within two-lane
4-minute averaging.  The depicted time interval has been extended to
make possible comparison of the congested state with the free flow. We
see that the flow rate at the detector D2 did not practically show any
visible dependence on the traffic state, whereas in the vicinity of
the detector D1 it exhibits strong variations rather than a
considerable fall. The latter feature prompts us that the congestion
observed near the detector D1 is not the developed ``stop-and-go
wave'' pattern. This statement, however, will be justified more
carefully below.

Let us now demonstrate that the two time series recorded at D1 and D2
present actually two synchronized modes different in nature. To
substantiate this statement we depicted in the row~3 left and right
windows the same time series of the vehicle velocity but on smaller
scales (from 10:00 to 11:00 a.m.) and averaged only over 30-second
interval. First, we see that the multilane correlations hold also on
these scales. Second, the difference between the two time series
becomes pronounced. The time series collected by D2 (the row~3 left
window) contains several quasi-plateaus in the time variations of the
vehicle velocities (singled out in this window and also in the row~4
left one by dotted ellipses). In contrast, the velocity time series
obtained by D1 seems to be no more than a collection of spikes varying
in amplitude. To argue for the statement about the different nature of
these congestions the row~4 left and right windows compare these
30-second averaging time series of the vehicle velocity and flow rate
simultaneously for the left lanes. For the detector D2 we meet a
visible distinction between the time pattern of vehicle velocity and
that of the flow rate. For the detector D1 they are quite similar in
appearance.

\begin{figure}
\begin{center}
\includegraphics[scale=1]{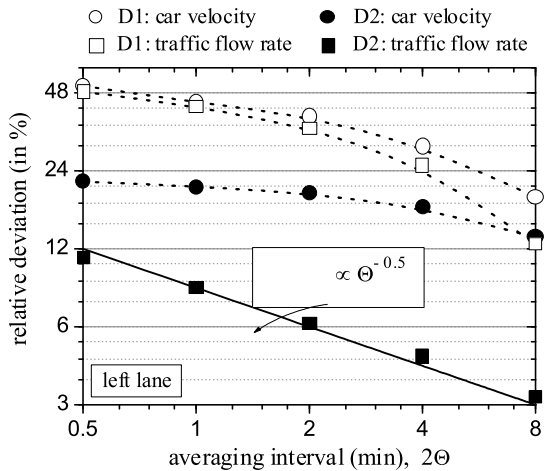}
\end{center}
\caption{%
  The relative deviations of the vehicle velocity and flow rate
  \textsl{vs} the averaging interval for the time series collected by
  the detectors D1 and D2 at the left lane on 6 June 1996 and 15 June
  1996, respectively, between 10 and 11 a.m.
  \label{fignew3}}
\end{figure}

To make this difference more pronounced we have analyzed the variance
for the time series of the vehicle velocity and flow rate collected by
the detectors D1 and D2 at the left lane during the observation time
from 10:00 to 11:00 a.m. Fig.~\ref{fignew3} presents the corresponding
relative deviations
$$
 \left.\frac{\delta v}{\left<v\right>}\right|_{\text{D1}} , \
 \left.\frac{\delta q}{\left<q\right>}\right|_{\text{D1}} , \
 \left.\frac{\delta v}{\left<v\right>}\right|_{\text{D2}} , \
 \left.\frac{\delta q}{\left<q\right>}\right|_{\text{D2}}
$$
depending on the averaging interval $2\Theta$. For the D1 time series
we meet at first a weak dependence of the relative deviations on the
averaging scale $2\Theta$ for both the vehicle velocity and flow rate
which, then, decrease fast keeping approximately their ratio fixed,
i.e.\
$$
 \left.\frac{\delta v(\Theta)}{\left<v\right>}\right|_{\text{D1}} \sim
 \left.\frac{\delta q(\Theta)}{\left<q\right>}\right|_{\text{D1}}.
$$
However, the question as to whether the latter estimate has a certain
physical meaning or just is a simple coincidence requires an
individual analysis. In contrast, the D2 time series demonstrates
another behavior. The relative deviation of the vehicle velocity
exhibits a weak decrease approximately from 22\% to 13\% as the
averaging interval grows 16-fold from 30 seconds to 8 minutes.
Conversely, the relative deviation of the flow rate is scaled
practically as (straight solid line in Fig.~\ref{fignew3})
\begin{equation}\label{1by2}
 \left.\frac{\delta q(\Theta)}{\left<q\right>}\right|_{\text{D2}}
 \approx 12\% \times \left(\frac{0.5\,\text{min}}{2\Theta}\right)^{\tfrac12}.
\end{equation}
Expression~(\ref{1by2}) enables us to suppose that the fluctuations of
the flow rate observed at the detector D2 during one hour from 10 to
11 a.m.\ are due to a certain white noise. Moreover, the mean flow
rate $\left.\left<q\right>\right|_{\text{D2, left lane}}$ during the
given time interval was about 1800~veh/h/l, so on average the detector
D2 recorded one car passing it on the left lane per two seconds.
Keeping in mind that the detectors measured the times of car passing
within seconds we may treat each record of a car having passed the
detector as a random event with the probability $p\approx 0.5$. Let us
consider these events mutually independent and describe them with a
sequence $\{\gamma(i)\}$ of random numbers such that take values equal
to either 1 or 0 with the probability $p$ and $1-p$, respective, i.e.
\begin{equation}\label{haha1}
    \left<\gamma(i)\right> = p\quad\text{and}\quad
    \left<\gamma(i)\gamma(i')\right> = p\delta_{ii'} + p^2(1-\delta_{ii'})\,,
\end{equation}
where $\delta_{ii'}$ is Kronecker's symbol. In these terms
Eq.~(\ref{qdef}) can be rewritten as
$$
  q_l(t,\Theta)  =  \sum_{i=-\infty}^{\infty} w_{\Theta}(t-i)\gamma(i)\,.
$$
Whence we immediately get the following expression for the mean
relative deviation of the random variable $q_l(t,\theta)$:
\begin{equation}\label{haha2}
  \frac{\left(\left<[q_l(t,\Theta) - \left<q_l(t,\Theta)\right>]^2\right>\right)^{\tfrac12}}
  {\left< q_l(t,\Theta) \right>} = \sqrt{\frac{1-p}{2p}\cdot\frac1{2\Theta}}\,.
\end{equation}
In obtaining Eq.~(\ref{haha2}) we have taken into account the
normalization condition~(\ref{ncond}) and for the sake of simplicity
extended the exponential dependence~(\ref{wcoef}) of the weight
coefficients $w_{\Theta}(t)$ to all the values of its argument $t$. It
should be recalled that the value $\Theta$ appearing in (\ref{wcoef})
matches the time scale measured in seconds.  For $p=0.5$ and the time
scale $\Theta$ measured in minutes Eq.~(\ref{haha2}) directly
gives us expression~(\ref{1by2}) with the replacement of 12\% by
13\%.

\begin{figure*}
\begin{center}
\includegraphics[scale=1]{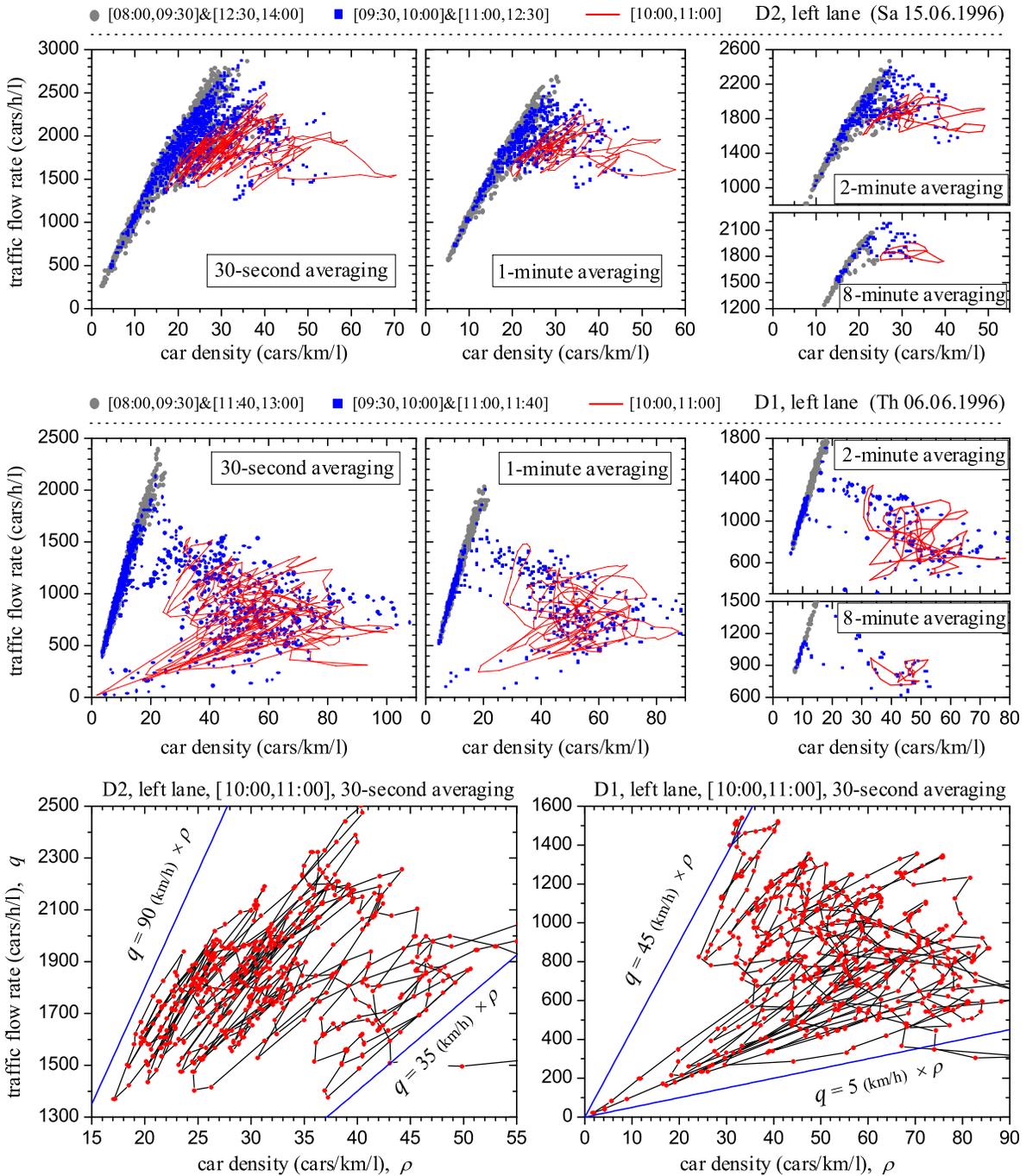}
\end{center}
\caption{%
  The fundamental diagrams showing the possible states of traffic flow
  in the $\rho q$-plane where for a chosen averaging interval
  $2\Theta$ the vehicle density $\rho(t,\Theta)$ was calculated as the
  ratio of the current flow rate $q(t,\Theta)$ to vehicle velocity
  $v(t,\Theta)$, i.e.\ $\rho = q/v$. The analyzed time series
  (visualized in Fig.~\protect\ref{fignew1}) and the corresponding
  colors are depicted at the window tops. Red color is used for the
  traffic flow states recorded within the time interval [10:00,11:00]
  a.m. (the corresponding time series are shown in
  Fig.~\protect\ref{fignew1}, the row~3,~4, left, right windows). Rows
  1 and 2 illustrate evolution of the fundamental diagrams depending
  on the averaging interval, whereas the windows in the third row show
  the [10:00,11:00] series in detail individually. In these windows
  the red dots are time series points connected by solid black lines
  according their order in the time series.  Blue lines are drawn to
  guide the eyes to follow the quasi-free traffic states specified by
  the relationship $q = v \rho$ between the flow rate $q$ and the vehicle
  density $\rho$ where the coefficient $v$ is a certain constant.
\label{fignew2}}
\end{figure*}

Thereby, the time variations in the flow rate recorded by the detector
D2 are likely to be due to the discreteness of traffic flow bearing no
correlations at all. In no case such fluctuations should be ascribed
to the state of traffic flow. Thus dealing with the congested traffic
in the vicinity of the detector D2 we actually meet the traffic flow
state where the vehicle velocity exhibits strong long-time
fluctuations with certain time plateaus and the traffic rate, in
contrast, has to be regarded as a constant value. These plateaus are
separated by sharp jumps as illustrated in Fig.~\ref{fignew1}, the
row~3 left and row~4 left windows with respect the plateaus singled
out by the first two dotted ellipses. In the row~3 left windows it is
also clearly visible that the plateau position with respect to the
$v$-axis varies substantially. It should be noted that Neubert
\textit{et al.} \cite{Neub99} also have found strong long-time
correlations in the vehicle velocity and their absence in the headway
distribution but with respect to the D1 time series. They also have
obtained the flow rate independence from the vehicle density for the
synchronized mode but after averaging over all the D1 time series.
Besides, there are cellular automata models predicting also a constant
value of the flow rate inside a certain interval of the vehicle
density which is due to local highway defects (see
Ref.~\onlinecite{CA1,CA2,CA3,CA4,CA5} and the review by Chowdhury
\textit{et al.} \cite{Sh00}). However, in the case under consideration
a constant value should be ascribed to the \textit{current} flow rate.
The latter enables us also to suppose that a possible spatial pattern
of the vehicle distribution was fixed on the highway during the
observation time. So the D2 time series of the vehicle velocity is
likely to visualize the real dynamics of traffic flow caused by its
internal local properties rather than by the pinch effects. In this
connection it should noted that Lee \textit{et al.}  \cite{L2,L3}
and Helbing \textit{et al.} \cite{Ha1} preferred to place the
congested traffic states matching spatial vehicle patterns fixed as a
whole on highways into an individual group.

In order to demonstrate that the analyzed traffic states belong in fact to the
synchronized mode we present the fundamental diagram (Fig.~\ref{fignew2})
reflecting the traffic flow states on the $\rho q$-plane. For a chosen
averaging interval $2\Theta$ the current values of the flow rate $q(t,\Theta)$
and the vehicle velocity $v(t,\Theta)$ were calculated according to
Eqs.~(\ref{qdef}), (\ref{vdef}), and then the effective vehicle density $\rho$
was calculated as $\rho(t,\Theta)=q(t,\Theta)/v(t,\Theta)$. The row~1 and
row~2 windows exhibit the fundamental diagrams for the traffic flow on the
left lane near the detectors D2 and D1, respectively, depending on the
averaging interval $2\Theta$. Red color has been used to make visible the
states of traffic flow recorded within the time from 10 a.m.\ to 11 a.m.\ and
thus to separate the free flow states from the those of the congested traffic.
We see that, in fact, the congested traffic flow under consideration possesses
the widely scattered state domain and so can be referred to as the
synchronized mode. It is most clear for the time series averaged over 1-minute
interval. On one hand, the widely scattered states cover a sufficiently large
domain on the $\rho q$-plane. On the other hand, for the D2 time series the
gap between this domain and the free flow region becomes quite visible. For
the D1 time series a certain ``tail'' of the widely scattered state domain in
the vicinity of the origin $(\rho=0,q=0)$ (see the row~2 left window) typical
for the ``stop-and-go way'' pattern~\cite{Neub99} disappears on these scales.
So the ``stop-and-go way'' traffic flow was not, at least, the dominant phase
state.

The windows in row~3 show the fundamental diagram for the traffic flow
states recorded during the time from 10 to 11 a.m. These windows
actually exhibit the structure of the widely scattered state domains
visualized in the given time series. Roughly speaking, both of them
comprise a large amount of straight lines describing quasi-free flow
states characterized by the relationship $q=\rho v$ with a constant
coefficient $v$ and the continuous transitions between them. The
particular magnitudes of $v$, however, do not coincide with the free
flow velocity being about 100 km/h (see Fig.~\ref{fignew1}, rows~1,2
right windows) but vary inside a wide interval from 35~km/h to 90~km/h
for the D2 series and from 5~km/h to 45~km/h for the D1 series.
Concerning the lower boundary of the latter interval, however, it
should be pointed out that the result can depend on the way of
calculating the vehicle density \cite{Neub99}. For the D2 time series
this feature remains also after averaging over 1-minute interval,
whereas for the D1 series it disappears and only the transition lines
are visible.

The row~1 and row~2 windows again raise the question about the
fundamental diagram which has been discussed in many papers (see,
e.g., reviews~\cite{Sh00,Hrev}), in particular, which averaging
interval should be used. As can be seen in these windows the scattered
state domain shrank considerably on scales about 8 minutes. Besides,
keeping in mind the drawn conclusion about the nature of the flow rate
fluctuations recorded by the detector D2 we can suppose that the
widely scattered states of the light synchronized mode are due to the
traffic flow discreteness. The vehicle velocity, however, exhibit
strong fluctuations comprising many long-lived states and the
continuous jump transitions between them.

\section{Synchronized mode as a glass-like phase state\label{sec:3}}

\begin{figure}
\begin{center}
\includegraphics[width=82mm]{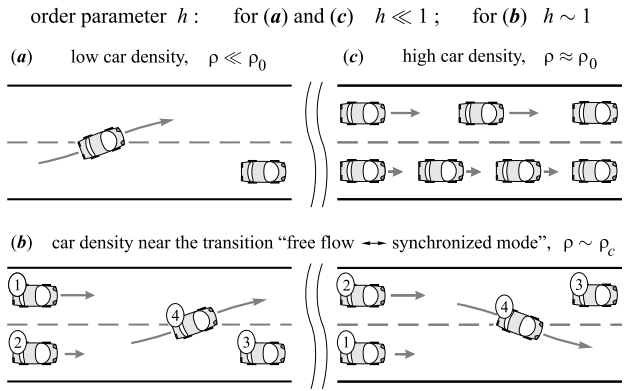}
\end{center}
\caption{Schematic illustration of the car arrangement in the various phases
  of traffic flow and the multilane vehicle interaction caused by cars
  changing the lanes.\label{Fold1}}
\end{figure}

When vehicles move on a multilane highway without changing the lanes
they interact practically with the nearest neighbors ahead only.
Therefore, there should be no internal correlations in the vehicle
flow at different lanes.
In particular, the drivers that would prefer to move faster than the
statistically mean driver will bunch up forming the platoons headed by
a relatively slower vehicle. When the cars begin to change lanes for
overtaking slow vehicles the car ensembles at different lanes will
affect one another. The cause of this interaction is due to the fact
that a car occupies during a lane change manoeuvre two lanes
simultaneously, affecting the cars moving behind it at both lanes.
Figure~\ref{Fold1}($b$) illustrates this interaction for cars 1 and 2
through car 4 changing the lanes.  The drivers of both cars 1 and 2
have to regard car 4 as the nearest neighbor and, so, their motion
will be correlated during the given manoeuvres. In the same way car 1
is affected by car 3 because the motion of car 4 directly depends on
the behavior of car 3. The more frequently lane changing is performed,
the more correlated traffic flow is on a multilane highway. Therefore,
to characterize traffic flow on multilane highways it is reasonable to
introduce an additional state variable, the order parameter $h$
\cite{W1,W2}. This variable is the mean density of such car triplets
normalized to the maximum possible for the given highway and, so, can
play the role of a measure of the multilane correlations in the
vehicle flow.

\begin{figure}[tbp]
\begin{center}
\includegraphics[width=82mm]{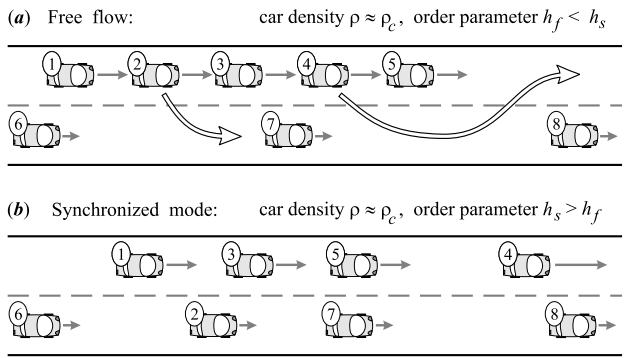}
\end{center}
\caption{Schematic illustration of the alteration in the vehicle arrangement
  near the ``free flow $\leftrightarrow $ synchronized mode'' phase
  transition.\label{FHPhT}}
\end{figure}

On the other hand, the order parameter $h$ introduced in this way can
be regarded as a measure of the vehicle arrangement regularity. Let us
discuss this question in detail for the free flow and synchronized
mode individually. In the free flow the feasibility of overtaking
makes the vehicle arrangement more regular because of platoon
dissipation. So, as the order parameter $h$ grows the free traffic
becomes more regular. Nevertheless, in this case the density of the
car multilane triplets remains relatively low, $h\ll 1$, and the
vehicle ensembles should exhibit weak correlations. Figure~\ref{FHPhT}
illustrates the $F\rightarrow S$ transition. As the car density grows
in free flow, the ``fast'' drivers that at first overtake slow
vehicles individually begin to gather into platoons headed by more
``slow'' cars among them but, nevertheless, moving faster than the
statistically mean vehicle, see Fig.~\ref{FHPhT}($a$). The platoons
are formed by drivers preferring to move as fast as possible keeping
short headway distance without lane changing. Such a state of the
traffic flow should be sufficiently inhomogeneous and is the reason
for the occurrence of high-flow states in free traffic. Therefore,
even at a sufficiently high car density the free flow should be
characterized by weak multilane correlations and not too big values of
the order parameter $h_{f}$. The structure of these platoons is also
inhomogeneous:
they comprise cars whose drivers would prefer to move at different
headways (for a fixed velocity) under comfortable conditions.
So, when the density of vehicles attains sufficiently high values and
their mean velocity decreases remarkably with respect to the velocity
on the empty highway some of the ``fast'' drivers can decide that
there is no reason to move so slowly at such short headway requiring
strain. Then they can either overtake the car heading the current
platoon by changing lanes individually or leave the platoon and take
vacant places, compare Fig.~\ref{FHPhT}($a$). The former has to
increase the multilane correlations and, in part, to decrease the mean
vehicle velocity because the other drivers should give place for this
manoeuvres in sufficiently dense traffic flow. The latter also will
decrease the mean vehicle velocity because these places are vacant
from the standpoint of sufficiently ``fast'' drivers only but not from
the point of view of the statistically mean ones preferring to keep
longer headway in comparison with the platoon headway. That means,
that the statistically mean drivers have to decelerate, decreasing the
mean vehicle velocity. The two manoeuvre types make the traffic flow
more homogeneous by dissipating the platoons and smoothing the headway
distribution, see Fig.~\ref{FHPhT}($b$). Besides, the single-vehicle
data \cite{Neub99} show that the synchronized mode is characterized by
long-distant correlations in the vehicle velocities.  The headway
fluctuations are correlated only on small scales. These findings
justify the assumptions of the synchronized mode being a more
homogeneous state than the free flow. We think \cite{W2} that the
given scenario describes the synchronized mode formation which must be
characterized by a large value of the order parameter, $h_{s}>h_{f}$,
and a lower velocity in comparison with the free flow at the same
vehicle density. It should be noted that this mechanism of the
synchronized mode emergence is related to the hypothesis by Kerner
\cite{KIsr,KL3,NP} that the synchronized mode is caused by a
``Z''-like form of the overtaking probability depending on the vehicle
density.

Keeping in mind this scenario we have assumed the mean velocity $v$ of
multilane traffic flow to be determined by both the vehicle density
$\rho $ and the order parameter $h$, namely, $v=\vartheta (h,\rho )$,
where the function $\vartheta (h,\rho )$ is regarded as a
phenomenological relationship known beforehand. Its general properties
have been discussed in Ref.~\onlinecite{W1,W2}, in particular, it has
been assumed to be a decreasing function with respect to $h$ for the
synchronized mode. Previously we have written the governing equation
for the order parameter completely describing the traffic state
evolution in terms of time variations in the mean velocity $v$ and the
order parameter $h$, the vehicle density is treated as a fixed value.
In this way we have obtained a simple and natural explanation of the
observed sequence of phase transitions $F\rightarrow S\rightarrow J$,
with each of them being of the first order. In particular, this model
predicts that the order parameter $h$ exhibits a sharp jump from the
value $h_{F}(\rho _{c})\ll 1$ corresponding to the free flow up to the
order parameter $h_{S}(\rho _{c})\lesssim 1$ of the synchronized mode
when the vehicle density $\rho $ exceeds a certain critical value
$\rho_{c}$. The further evolution of the synchronized mode as the
vehicle density grows is given via the dependence
$v=\vartheta[h_{S}(\rho),\rho ]$.

\begin{figure}[tbp]
\begin{center}
\includegraphics[width=82mm]{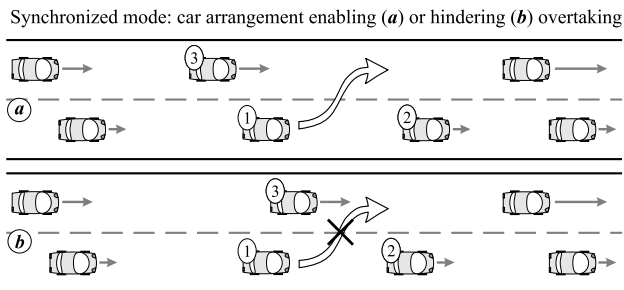}
\end{center}
\caption{Schematic illustration of the car arrangement at the
  neighboring lane in the synchronized mode that enables overtaking
  ($a$) or hinders it ($b$).\label{Fold3}}
\end{figure}

However, to describe the widely scattered state region of the
synchronized mode we have to regard the multilane car interaction into
more detail. The matter is that for a car to be able to leave a
platoon or to change a lane, the local vehicle arrangement at the
neighboring lanes should be of a special form (Fig.~\ref{Fold3}). So
the car rearrangement essentially depends also on the particular
details of the neighboring car configuration exhibiting substantial
fluctuations. These fluctuations, however, are of another nature than
those in the free flow. The latter can be treated as a white noise and
slightly blurs the dependence $v=\vartheta (h,\rho )$ for the free
flow (Fig.~\ref{exp}). In the synchronized mode, by contrast,
attaining the optimal conditions of driving, including also overtaking
slow vehicles, can be frustrated for a certain time as illustrated in
Fig.~\ref{Fold3}. For car~1 to be able to overtake car~2 the
neighboring car~3 should provide a room for this manoeuvre. Otherwise
the driver of car~1 has to wait and the local car arrangement will not
vary substantially. In other words, changes in the particular
realizations of the local car arrangement can be frozen for a certain
time although the globally optimal car configuration is not attained
at the current time moment. Due to this self-freezing effect the
synchronized mode can comprise a great amount of locally metastable
states and correspond to a certain two-dimensional region in the $\rho
q$-plane rather than to a line $q=\vartheta (\rho )\rho $. This
feature seems to be similar to that met in physical media with local
order, for example, in glasses where phase transitions are
characterized by wide range of controlling parameters (temperature,
pressure, \textit{etc.}) rather than their fixed values (see, e.g.,
\cite{Ziman}).

The synchronized mode consists of clusters of some ten cars moving
along the road as a whole \cite{Neub99} which will be called for
convenience the fundamental clusters of the synchronized mode.
Thereby, the introduced order parameter $h$ is actually an averaged
characteristics of such a cluster as whole and, so, it does not allow
for local fluctuations in the car arrangement inside this cluster. To
describe the latter another state variable, the order parameter $a$
has to be introduced. Its physical meaning is to allow for the fact
that the optimal and comfortable way of driving which the drivers try
to attain individually depends essentially on the current realization
of the car arrangement inside the given fundamental cluster. In other
words, we assume that the order parameter $h_{0}(a,\rho )$ matching
this optimal way of driving is a function of the vehicle density
$\rho$ as well as the order parameter $a$. Without lost of generality
in such a phenomenological approach we may relate $h_{0}(a=0,\rho )$
to the global optimum of the driving manner attained after averaging
over all the realizations of the car configuration, i.e.\ set
$h_{0}(a=0,\rho )=h_{S}(\rho )$. Besides, it is reasonable to relate
variations of the order parameter $a$ of amplitude about unity,
$\left\langle a^{2}\right\rangle ^{1/2} \sim 1$, with all the possible
realizations of the car configuration.

To complete the description of the long-lived state continuum of the
synchronized mode we should specify the evolution of the order
parameter $a$. Since the order parameter $a$ allows for microscopic
details of the fundamental cluster structure its fluctuations will be
treated as a random noise whose amplitude depend on the vehicle
density only. In contrast, the rate $da/dt$ of time variations in the
order parameter $a$ has to be affected substantially by the current
value of the order parameter $h$. In fact, as the order parameter $h$
tends to the local optimum value $h_{0}(a,\rho )$ for the given $a$
the rate $da/dt$ should be depressed because all the drivers forming
the fundamental cluster prefer to wait until a more comfortable car
configuration arises therefore inhibiting the evolution of the
fundamental cluster structure.

Now, the model can be formulated. However, before doing this it is
worth noting that the present paper does not describes the formation
and evolution of the synchronized mode \textit{inside} the free flow.
We construct a continuum model for the long-lived states. So we may
confine our consideration solely to the local properties ignoring the
interaction between fragments of the synchronized traffic flow
separated in space. The latter is worthy of individual and detailed
investigations.

\section{The dynamical trap model\label{sec:4}}

Keeping in mind the aforesaid and following \cite{W1,W2}, we specify
the local state of traffic flow on a multilane highway by the system
of four phase variables, the mean velocity $v$, the density $\rho $ of
vehicles, and two order parameters $h$ and $a$. The parameter $h$
describes the effect of multilane car interaction on the individual
driver behavior, so, the mean velocity $v=\vartheta (h,\rho )$ is a
direct function of the order parameter $h$ and the vehicle density
$\rho$ given beforehand and meeting general properties described in
\cite{W1,W2}. The other order parameter $a$ allows for fluctuations in
the car arrangement inside a fundamental cluster of the synchronized
mode and affects the relaxation process of the order parameter $h$,
namely, we write
\begin{equation}
\tau \frac{d\,h}{dt}=-\phi (h,a,\rho )\,+\epsilon \sqrt{\tau }\xi (t) .
\label{e.1}
\end{equation}
Here, $\tau $ is the average time drivers require to come to the
decision to begin or to stop overtaking manoeuvres, the function
$\phi(h,a,\rho )$ describes the regular behavior of individual drivers
depending on the traffic state as well as on the particular details of
the local car arrangement. Finally, the term $\epsilon \sqrt{\tau
  }\xi(t)$, a stochastic Langevin force, allows for a random component
in the driver evaluation of the current situation on the road:
\begin{equation}
\aver{\xi(t)} = 0\,,\quad
\aver{\xi(t) \, \xi(t^{\prime })} = 2\delta (t-t^{\prime })\, .  \label{e.2}
\end{equation}
The factor $\epsilon$ quantifies the size of the noise. As discussed
in the previous section it is reasonable to normalize the order
parameters $h$ and $a$ to their maximum values, which enables us to
accept that in the free flow the averaged value of the order parameter
$h_{F}[\rho ]$ is much less than unity, $h_{F}[\rho ]\ll 1$, whereas
in the synchronized mode it is about one, $h_{S}[\rho ] \lesssim 1$,
and to assume that the order parameter $a$ varies around zero, i.e.\
$\aver{a^2} \sim 1$.

In order to proceed, recall briefly the key features of the
synchronized mode.  When the vehicle density exceeds a certain
critical value $\rho _{c}$ the free flow becomes unstable and the
synchronized modes arises on the road through a first order
transition.  Our previous model describes this $F \rightarrow S$
transition as a sharp jump in the order parameter $h$ from $h_{F}[\rho
]$ to $h_{S}[\rho ]$ at a fixed value of the vehicle density $\rho $
which should be accompanied with stepwise jumps in the mean velocity
$\delta v=v_{F}[\rho ]-v_{S}[\rho ]$ and the traffic flow rate $\delta
q=$ $\rho \cdot \delta v$.  These jumps stem from the dependence of
the steady state order parameter $h_{\text{st}}[\rho ]$ on the vehicle
density $\rho $ taking the multivalued form for $ \rho >\rho _{c}$.
In the given case the points $(\rho ,q_{F}[\rho ])$ and $ (\rho
,q_{S}[\rho ])$ on the $\rho q$-plane have to correspond to well
determined states of traffic flow. However, the available empirical
data (illustrated in Fig.~\ref{exp}) justify the latter assumption
with respect to the free flow only, whereas the synchronized mode
stretches over a whole region with, may be, blurred boundaries
$q_{S}^{(l)}[\rho ]$ and $ q_{S}^{(u)}[\rho ]$. Besides, this region
turns out to be sufficiently thick, $\left( q_{S}^{(u)}[\rho
  ]-q_{S}^{(l)}[\rho ]\right) \sim q_{F}[\rho _{c}]$ and adjacent to
the free flow branch, $\left( q_{F}[\rho ]-q_{S}^{(u)}[\rho ]\right)
\ll q_{F}[\rho _{c}]$. We intend to explain the existence of the
synchronized mode continuum by the effect of random fluctuations in
the vehicle velocity $v$. However, it is not sufficient to introduce
additional stochastic Langevin forces into governing equations, as
done with respect to equation~(\ref{e.1}). Indeed, in this case such a
stochastic force quantifies the random component of the
\textit{individual} driver behavior, so, it should demonstrate similar
features in both the free flow and synchronized mode. Therefore, in
particular, we consider the amplitude $\epsilon $ of the introduced
stochastic force as well as its correlation function~(\ref{e.2}) to be
independent of traffic flow state. The free flow matches a line on the
$\rho q$-plane, the free flow branch, which is slightly blurred due to
random fluctuations. Besides, it exhibits no long-time correlation.
Thereby, an additional Langevin force cannot explain the wide
quasi-equilibrium state continuum of the synchronized mode. Moreover,
these Langevin forces cannot be strong, so, we may set $\epsilon \ll
1$ in equation~(\ref{e.1}).

Furthermore, we assume that in the synchronized mode where the multilane car
interaction is essential the order parameter $h$ is substantially affected by
the order parameter $a$ quantifying the random fluctuations in the local car
arrangement. In this case the possibility of a self-freezing effect discussed
above is responsible for the emergence of the long-lived state continuum.
Therefore, the next step should specify the function $\phi (h,a,\rho )$ in
such a way that the resulting steady-state dependence $h_{\text{st}}(\rho ,a)$
on the order parameter $a$ is strong. Since we deal solely with the
synchronized mode itself we can write down
\begin{equation}
\phi (h,a,\rho )=h-h_{0}(a,\rho )  \label{e.3}
\end{equation}
without lost of generality. Here the function $h_{0}(a,\rho )$ can be
approximated by the expression:
\begin{equation}
h_{0}(a,\rho )=h_{S}(\rho )-\Delta a^{2}  \label{e.4}
\end{equation}
which assumes that a zero value of the order parameter $a$ corresponds
to the most comfortable conditions for driving, to the highest order
of the traffic flow, and to the slowest vehicle motion. We note that
in the previous model~\cite{W1,W2} the value $h_{S}(\rho )$
characterizes the stationary state of the synchronized mode at the
given vehicle density $\rho$. The deviation of the car arrangement
from the optimal configuration destroys the order of traffic flow,
bringing the synchronized mode closer to the free flow in properties.
Therefore, we may set $\Delta $ approximately equal to unity, $\Delta
\sim 1$, and to regard the traffic state with $h\approx h_{0}(1,\rho
)$ as practically the free flow or, more rigorously, the state
directly adjacent to the free flow.

Now let us specify the governing equation for the order parameter $a$.
Local variations of the car arrangement seem to be determined solely
by the vehicle density. Indeed, the order parameter $h$ is actually
produced by averaging, first, over many particular realizations of the
fundamental cluster structure similar in properties with respect to
the possibility of the lane changing and, second, over all the cars
making up the fundamental cluster. So it is sufficiently rough
characteristics of the traffic flow state and the dynamics of the
order parameter $a$ can be treated as an independent of $h$ random
process of unit amplitude and induced by the white noise except for
the rate $\Omega (h,a)$ at which the parameter $a$ responds to the
white noise. Namely, we write
\begin{equation}
\tau \frac{da}{dt}=-\Omega (h,a)a+\Omega ^{1/2}(h,a)\sqrt{\tau }\varsigma
(t)\,,  \label{e.5}
\end{equation}
where again the noise $\varsigma (t)$ meets the conditions
\begin{equation}
\aver{\varsigma(t)} = 0, \ \left\langle \varsigma (t)\xi (t^{\prime
})\right\rangle =0,\ \left\langle \varsigma (t)\varsigma (t^{\prime
})\right\rangle =2\delta (t-t^{\prime })  \label{e.7}
\end{equation}
and the factor $\Omega (h,a)$ describes the self-freezing effect of
fluctuations in the car arrangement. In particular, when the order
parameter comes close to the local equilibrium, $h\rightarrow
h_{0}(a,\rho )$, i.e.\ $ \phi (h,a,\rho )\rightarrow 0$ the rate
$\Omega (h,a)$ should tend to zero. Therefore, by keeping only the
leading term we get
\begin{equation}
\Omega (h,a)\approx \left\{
\begin{array}{ccc}
\phi ^{2}(h,a,\rho )/\phi _{0}^{2} & \text{if} & \left| \phi (h,a,\rho
)\right| \ll \phi _{0} \\
\Omega _{0} & \text{if} & \left| \phi (h,a,\rho )\right| \gtrsim \phi _{0}
\end{array}
\right. \,.\text{\thinspace }  \label{e.8}
\end{equation}
Here, the constant coefficients $\phi _{0}\sim \Omega _{0}\sim 1$
because under the general conditions time variations of the behavior
of individual drivers as well as of the car configuration of
fundamental cluster should proceed essentially at the same rate. The
form of the random force in equation~(\ref{e.5}) has been chosen so
that the amplitude of fluctuations in variable $a$ remains unchanged
provided the factor $\Omega (h,a)$ is constant.  In other words, the
proximity of the fundamental cluster structure to a local equilibrium
slows down the time variations in the order parameter $a$. In some
sense the curve $h=h_{0}(a,\rho )$ on the $ah$-plane (for a fixed
value of $\rho $) forms certain traps for the path $ \{a(t),h(t)\}$,
which has given the model its name. If the order parameter $h$ had
been constant, governing equation~(\ref{e.5}) would describe the
collapse of the random motion $\{a(t)\}$, with the Ito process
corresponding to the maximum of the collapse intensity (see, e.g.,
Ref.~\onlinecite{Coll}). Therefore we also ascribe to the stochastic
equation~(\ref{e.5}) the Ito type, which completes the model
formulation. Naturally, in the given model the vehicle density $\rho $
is a fixed constant. Time variations in the order parameter $h(t)$ can
be mapped onto the evolution of the traffic flow rate $q(t)=\rho
\vartheta (h(t),\rho )$. Therefore, in the following it is sufficient
to regard the order parameters $a$ and $h$ only.

Here we present only the preliminary investigation of the stated
model, namely, the results obtained by its numerical simulation, which
demonstrate that this model in fact leads the desired long-lived state
continuum. For the sake of continuance in numerical simulations we
have converted to the dimensionless time $t\rightarrow t/\tau $,
introduced the variable $\eta =(h_{S}(\rho )-h)/\Delta $, renormalized
the constants $\epsilon \rightarrow \epsilon /\Delta $ and
$\phi_{0}\rightarrow \phi _{0}\Delta $, set $\Omega _{0}=1$. This
leads to:
\begin{eqnarray}
\frac{d\eta }{dt} &=&-(\eta -a^{2})+\epsilon \xi (t)  \label{e.10} \\
\frac{da}{dt} &=&-\Omega _{0}\varpi (\eta ,a)a+\Omega _{0}^{1/2}\varpi
^{1/2}(\eta ,a)\varsigma (t)\,  \label{e.11}
\end{eqnarray}
where the noise $\xi (t)$ and $\varsigma (t)$ meets the
conditions~(\ref{e.2}) and (\ref{e.7}). Finally, the function
$\varpi(\eta ,a)$ may be specified as:
\begin{equation*}
\varpi (\eta ,a)=\left\{
\begin{array}{ccc}
(\eta -a^{2})\,^{2}/\phi _{0}^{2} & \text{if} & \left| \eta -a^{2}\right|
\leq \phi _{0} \\
1 & \text{if} & \left| \eta -a^{2}\right| >\phi _{0}
\end{array}
\right. \,.
\end{equation*}
Again, the stochastic equation~(\ref{e.11}) is assumed to be of the
Ito type.

\begin{figure}
\begin{center}
\includegraphics{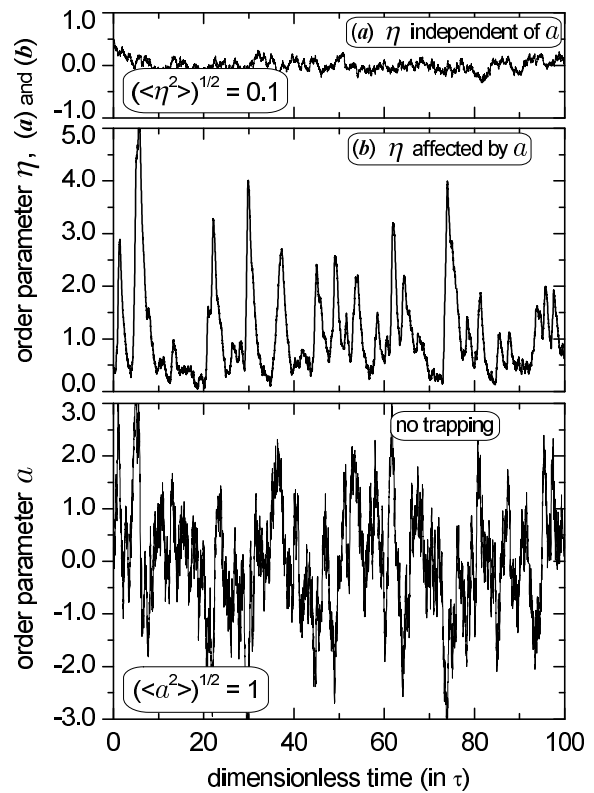}
\end{center}
\caption{The time pattern of the order parameters $\eta$ and $a$ if
  the self-freezing effect is suppressed. The window ($a$) exhibits
  time variations in the order parameter $\eta$ in the case where the
  influence of the order parameter $a$ has been ignored at all. The
  window ($b$) presents the dynamics of $\eta$ affected by
  $a$.\label{F1}}
\end{figure}

Let us now discuss the results obtained by simulating numerically the
system of Eqs.~(\ref{e.10}), (\ref{e.11}) for $\Omega _{0}=1$,
$\epsilon=0.1$, and $\phi _{0}=0.5$. First of all, Fig.~\ref{F1}
illustrates the evolution of the order parameters $\eta$ and $a$ when
there is no self-freezing.  In this case the order parameter $a$
exhibits the standard random pattern of Brownian movement inside a
region of unit width. We see a collection of practically independent
spikes of unit width. A similar pattern (see Fig.~\ref{F1}($a$)) is
demonstrated by the dynamics of the order parameter $\eta$ provided
the interaction of the parameters $\eta$ and $a$ have been ignored,
i.e.\ by replacing $(\eta -a^{2})$ by $\eta$. Naturally, in this case
the synchronized mode matches a line on the $\rho q$-plane for
$\epsilon \ll 1$. The dynamics of the order parameter $\eta$ affected
by the variable $a$ but without the reciprocal influence is shown in
Fig.~\ref{F1}($b$). Again a collection of spikes can be seen, whose
amplitude as well as width has increased tenfold. In other words, if
we take into account the effect of the fluctuations in the car
arrangement (the order parameter $a$) on the individual driver
behavior (the order parameter $h$) then it is possible to explain an
essential blurring of the synchronized mode state.

\begin{figure}
\begin{center}
\includegraphics{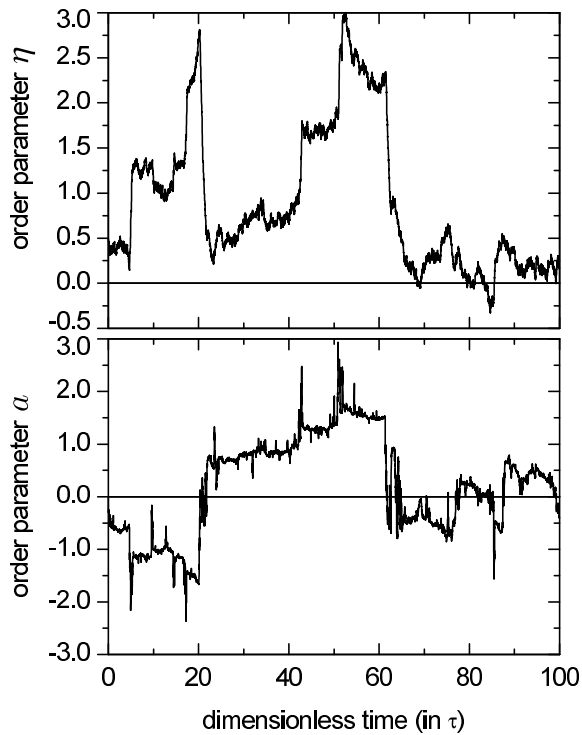}
\end{center}
\caption{The time pattern of the order parameters $\eta$ and $a$ when the
  self-freezing effect is substantial.\label{F2}}
\end{figure}

The dynamics changes dramatically for the full problem, see
Fig.~\ref{F2}. The time pattern takes a form corresponding to the
long-lived state continuum.  When the point $\{a(t),\eta(t)\}$
representing the current state of the synchronized mode wanders on the
$a \eta$-plane and reaches the curve $\eta =a^{2}$ at any point it
will be trapped for a certain time until it finally escapes from the
trap due to the noise $\epsilon\xi(t)$.  After that the system again
wanders in the $a \eta $-plane during a time interval about unity
before being trapped for the next time. Since the characteristic
duration of the trapping is much longer than unity the pattern looks
like a certain collection of local metastable states of the
synchronized mode.  However, such prolonged stays of the system are
not metastable in the rigorous meaning and we preferred to call them
simply the long-lived states. Since each point of the curve
$\eta=a^{2}$ is a trap, the long-lived states make up a certain
continuum.

\section{Remarks about governing equations for light synchronized traffic}

In writing equations~(\ref{e.10}) and (\ref{e.11}) we actually have
considered the state of the traffic flow homogeneous. Thereby, the
interactions between different fragments separated in space have been
neglected. In order to analyze a more realistic case of the evolution
of traffic flow with the long-lived states we have to deal with the
fields $\rho(x,t)$, $\eta(x,t)$, and $a(x,t)$ describing the vehicle
distribution along the highway, the order of the car arrangement as
well as local fluctuations in the car arrangement at different highway
parts. It should be noted that the given model has been proposed for
the light synchronized mode where the driver's delay to the headway
variations seems to be of minor importance. This enables us to assume
the local value of the vehicle velocity practically equal to the
optimal velocity $\vartheta(h,\rho)$ for current values of the vehicle
density $\rho$ and the order parameter $h$.

In this section we only touch a possible way of constructing the
distributed model for traffic flow with long-lived states. For
inhomogeneous traffic flow Eqs.~(\ref{e.10}), (\ref{e.11}) actually
describe the process in the frame locally ``attached'' to the traffic
flow, i.e.\ moving at the same speed along the road. So, in order to
return to the physical coordinate system we, first, should replace the
time derivative by the material derivative, i.e.
$$
\frac{d}{dt} \Rightarrow \partial_t + u(\eta,\rho)\partial_x\,,
$$
where we again use the dimensionless form of time $t$ in the term
$u(\eta,\rho)=\tau\vartheta[(h_S(\rho)-\Delta\eta),\rho]$.

Second, as was discussed in Sec.~\ref{sec:3} the order parameter $h$
characterizes the arrangement of the fundamental car cluster of
synchronized traffic flow. So it cannot exhibit considerable
variations on spatial scales about the characteristics size $\ell$ of
this cluster. So the right-hand side of equation~(\ref{e.10}) has to
be modified to allow for this effect. Following our previous
paper~\cite{W2} we introduce in equation~(\ref{e.10}) the term
$$
\ell^2\partial_x^2 h + 2^{-1/2}\ell\partial_x h\,,
$$
smoothing the field $h$ on the scales $\ell$ and taking into account
the asymmetry in the vehicles interaction. Conversely, the order
parameter $a$ allows for local variations of the car arrangement
inside the fundamental car cluster, so we may regard the field
$a(x,t)$ as totally independent at different points along highway, at
least, within such a mesoscopic description.

In this way the system of equations~(\ref{e.10}) and (\ref{e.11}) can
be extended as follows:
\begin{eqnarray}
 \nonumber
\partial_t\eta + u(\eta,\rho)\partial_x\eta &=&  \ell^2\partial_x^2 h
+ 2^{-1/2}\ell\partial_x h \\
&&\quad {} -(\eta -a^{2}) + \epsilon \xi (t,x)
 \label{e.r1} \\
\partial_t a + u(\eta,\rho)\partial_x a  &=& -\Omega _{0}\varpi (\eta
,a)a\nonumber\\
&&\quad {} + \Omega _{0}^{1/2}\varpi^{1/2}(\eta ,a)\varsigma (t,x)\,,
 \label{e.r2}
\end{eqnarray}
where the terms $\xi(t,x)$ and $\varsigma(t,x)$ are random sources
uncorrelated in time and space. Naturally, these equations must be
completed by the continuity equation:
\begin{equation}\label{e.r0}
\partial_t \rho + \partial_x\left[u(\eta,\rho)\rho \right]  =
(D\tau)\partial_x^2\rho\,,
\end{equation}
where $D$ is the effective diffusion coefficient.

The proposed distributed model for the light synchronized traffic flow
is worthy of an individual investigation. Here we have only
demonstrated the feasibility of its construction.

\section{Conclusion}

We proposed a novel notion of the widely scattered states in the
synchronized traffic stimulated by Kerner's
hypotheses~\cite{K2,KTGF99,KIsr,KL3,H1H2} about the existence of a
multitude of metastable states in the homogeneous or
``homogeneous-in-speed'' synchronized traffic.

First, we have analyzed the single-vehicle data collected at the
German highway A1 near Colonge from 6-th to 17-th of June 1996 in
order to verify the presence of the hypothetical multitude of
metastable states. As a preliminary result we have found two types of
synchronized mode observed at different detectors. One of them can be
classified as the ``oscillating congested traffic'' \cite{Hrev} and
has been analyzed in detail in \cite{Neub99}. The behavior of the
other synchronized mode differs essentially therefore we preferred to
classify it as an individual type called light synchronized traffic.
Exactly this type of traffic flow possesses the hypothetical multitude
of quasi-stationary states. Namely, it has been found that the vehicle
velocity dynamics consists of a collection of certain fragments inside
which the vehicle velocity is approximately constant and the
continuous transitions between them occur via sharp jumps. The
particular values of the vehicle velocity in these fragments vary in a
wide interval. In contrast, the traffic flow rate has to be regarded
as a constant because, as we have shown, its fluctuations are mainly
due to the traffic flow discreteness. In spite of the latter
properties the fundamental diagram drawn for this traffic mode in the
standard way contains the widely scattered state domain. Keeping in
mind the obtained results we preferred to call this domain the
long-lived state region, because it is made up of states that are not
stationary at all, at least in the standard physical meaning.

Second, we have proposed a simple mathematical model explaining this
complex behavior of light synchronized traffic in terms of dynamical
traps. Advancing the idea proposed in~\cite{W1,W2}, we specify the
state of traffic flow by four phase variables, the mean velocity $v$,
the vehicle density $\rho$, and two order parameters $h$ and $a$. The
parameter $h$ describes the effect of multilane correlations in the
vehicle motion on the behavior of individual drivers originating from
lane change manoeuvres. On the other hand, it characterizes the order
of traffic flow, so the transition from the free flow to the
synchronized mode is represented as a sharp jump in the value of the
order parameter $h$.  We assume the existence of the relationship
between the mean velocity $v$, the vehicle density $\rho$ and the
order parameter $h$, i.e.\ $v=\vartheta(h,\rho)$, where
$\vartheta(h,\rho)$ is a certain phenomenological function known
beforehand. In these terms we can describe the local dynamics of the
synchronized mode in terms of the evolution of the order parameter
$h$, here the vehicle density $\rho$ is treated as a fixed constant.
The latter is justified because the synchronized mode is characterized
by long-distance correlations, so its basic features should stem from
the main properties of a certain fundamental cluster, the minimal
fragment of car ensemble that moves as a whole and can be regarded as
an atomic element of the synchronized mode.

The dynamics of the order parameter $h$ is governed by the vehicle
density and in addition, by local fluctuations in the car arrangement
affecting the feasibility of changing lanes for overtaking. The latter
aspect is the key point of the mechanism responsible for the emergence
of widely scattered synchronized states. Due to different realizations
of the car arrangement inside the fundamental cluster there exist a
continuum of locally quasi-equilibrium states of the order parameter
$h$. So, when such a local quasi-equilibrium state is attained, the
individual drivers prefer to wait for a more comfortable car
configuration to continue overtaking. In this case also the rate of
changes in the car arrangement slows down and, as a result, the
current state of traffic flow is `frozen'. Exactly this effect of
self-freezing gives rise to a continuum of long-lived states in the
light synchronized traffic.

We have written a particular mathematical description of self-freezing
and justified these qualitative conclusions by solving the stated
model numerically.

\begin{acknowledgments}
  The authors would like to thank the Ministry of Economy and
  Transport, and the Landschaftsverband Rheinland in the German state
  Northrhine-Westfalia for the support in obtaining this beautiful
  data-set. Also
  this work has been supported in part by Russian Foundation of Basic Research,
  Grant \#~01-01-00389.
\end{acknowledgments}

\end{document}